# Optical properties of spin coated and sol-gel dip coated cupric oxide thin films

P. Samarasekara and N.G.K.V.M. Premasiri

Department of Physics, University of Peradeniya, Peradeniya, Sri Lanka

## Abstract

Spin coating technique was employed to fabricate CuO films at different spin speeds for different time duration, and they were annealed at different temperatures for different time durations in air. These thin films were characterized using UV/Vis spectrometer and solar cell simulator. As the spinning rate (rpm) increases, the corresponding band gap energy of the thin film increases. However, the annealing temperature does not affect the optical band gap. Moreover, the optical band gap doesn't depend on the annealing time and spin time. It was observed that the optical band gap decreases with the number of CuO layers. The optical band gap values of CuO thin films were between 1.843 and 1.869 eV. Inclusion of the two additives enhances the photocurrent, photovoltage and efficiency. Photocurrent density can be enhanced from 0.1464 to 1.172 mA cm$^{-2}$ using the additive ethylene glycol. Properties of CuO films synthesized using sol-gel dip coating technique were compared with those of CuO films grown using spin coating method.

## 1. Introduction:

Cupric oxide (CuO) is a black color compound with slight translucence. It has a monoclinic crystal structure. CuO is a prime candidate in the solar energy applications, diode fabrications, lithium ion batteries, electrochemical cells, superconductors [1] photovoltaic [2], nanoelectronics [3] and spintronics [4]. It is a p-type semiconductor with band gap of 1.2-1.9eV [2]. Thin films of CuO have been synthesized using sol-gel spin coating [2,5], electrodeposition [6], spray pyrolysis technique [7]. Variation of conductivity and photoconductivity of CuO thin films have been studied as a function of layers [2]. Cuprous oxide thin films were fabricated on silicon substrate using the sol-gel spin coating method [5]. The band gaps of those films were recorded to be on the range of 2.0-2.2 eV. Effect of deposited potential on the physical properties of electrodeposited CuO thin films has explored [6]. CuO thin films prepared by spray pyrolysis technique were characterized using structural, electrical and optical properties [7]. XRD, absorption spectroscopy and conductivity/ resistivity analyses were used. 350 °C was used as the substrate temperature. A Hall-effect measurement system was utilized to measure Hall-coefficient, charge carrier concentration, mobility, resistivity and conductivity of the films [7]. Among many other deposition techniques, spin coating method was found to be low cost, reliable and fast.

CuO exhibits Room Temperature Ferromagnetism [4], and thus this material offers a very good option for a class of spintronics, especially without the presence of any transition metal. Second and third order perturbed Heisenberg Hamiltonian was employed to investigate the magnetic properties of ferrite and ferromagnetic films by us [10, 14-18]. Thin films of multi walled carbon nanotubes [8] and $Cu_2O$/CuO layers [9] have been fabricated. In this manuscript, the optical properties of spin coated CuO films have been described. Copper oxide was fabricated using reactive dc sputtering by us [11]. Energy gap of semiconductor particles doped with salts were determined by us [12]. ZnO films were synthesized by reactive DC sputtering [13].

## 2. Experimental:

First diethanolamine, cupric acetate and isopropyl alcohol were stirred together. The mole ratio between cupric acetate and diethanolamine was 1:1. Thereafter, a solution with $Cu^{+2}$ concentration of 1.5 mol $dm^{-3}$ was prepared with isopropyl alcohol as the solvent. Then, this solution was continuously stirred for 24 hours. This stirred solution was used for sol-gel spin coating. In addition to the basic solution of $Cu^{+2}$, 6 different solutions were prepared using those two additives; 3 solutions per each additive, varying weight percentage. For this purpose, the additives were included in the solution at the initial step of the experimental procedure. Films were synthesized at three different spin rates (1500, 2200 and 2400 rpm), and two spin times (15 and 60 sec.). Spin coated thin films were subsequently annealed in air at temperatures of 150, 250, 350, 450 and 550 °C for 1 and 2 hours.

Some samples were synthesized using sol-gel dip coating method. For this purpose, a glass substrate was dipped in a solution with the same composition of cupric acetate, diethanolamine and isopropyl alcohol as used for the spin coating. The mole ratio between cupric acetate and diethanolamine was used to be 1:1. Then, a solution with $Cu^{+2}$ concentration of 1.5 mol $dm^{-3}$ was prepared, using isopropyl alcohol as the solvent. Initially, the glass substrate was dipped in the solution for 24 hours, and the resulting thin film had not acquired the adequate evenness. Then, the dipped time was extended to 48 hours where the thin film was rather uniform.

Structural properties of films were investigated using X-ray diffraction (XRD) method with Cu-Kα radiation. Agilent UV-Visible Spectrophotometer (8453 UV-Vis) was used to measure UV/visible spectroscopy of samples. Initially, a background analysis was performed using a cleaned piece of glass, and then thin film sample was analyzed for UV/Vis spectrum. In order to analyze thin films for their photovoltaic characteristics, CuO thin films were spin-coated on Fluorine-doped Tin Oxide (FTO) coated glass slides; Thickness = 3 mm; Surface resistivity ~10 Ω/sq]. Then, these films were studied using a solar cell simulation system (Portable Solar Simulator PEC-L01; Peccell Technologies, Inc.).

## 3. Results and discussion:

Figure 1 shows the XRD pattern of CuO film spin coated at 1500 rpm for 15s. This film was subsequently annealed in air at 450 $^0$C for 1 hour. XRD pattern indicates the formation of the CuO phase. Peaks belonging to any secondary phases are not found in XRD pattern. CuO could be crystallized only above the annealing temperature of 350 $^0$C. Below 350 $^0$C, samples were amorphous, and XRD patterns didn't indicate any peaks.

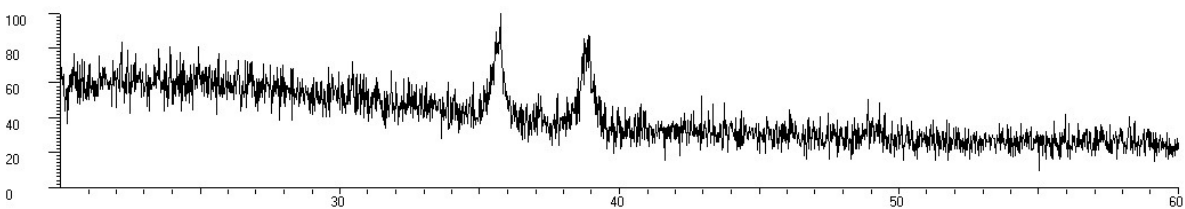

Figure 1: XRD pattern of CuO annealed at 450 $^0$C for 1 hour.

UV/Vis spectroscopy was essentially used in determining the optical band gap of cupric oxide thin films. Thereupon, the response of the optical band gap to the variations in the chemical composition, spin-coating conditions and annealing conditions were studied. By measuring the absorption (A) of CuO film at different wavelengths using UV/Vis spectrometer, the absorption coefficient (α) at each frequency (ν) was calculated. The method of Tauc plots was used to investigate the optical band gap. The point, at which the tangential line drawn to the curve intercepts the horizontal axis of graph, gives the optical band gap. Figure 2 shows the Tauc plot of $(\alpha h\nu/A)^2$ vs. $h\nu$ (Photon Energy) for a thin film spin coated at 1500 rpm for 15 s and annealed at 350 °C for 1 hour. Here h is the Planck's constant. Figure 3 indicates the Tauc plot of $(\alpha h\nu/A)^2$ vs. $h\nu$ for a thin film spin coated at 1500rpm for 60 s and annealed at 450 °C for 2 hours. Figure 4 shows the Tauc plot of $(\alpha h\nu/A)^2$ vs. $h\nu$ for a thin film spin coated at 2200 rpm for 15 s and annealed at 550 °C for 1 hour.

The optical band gap values calculated from Tauc plots are tabulated in table 1. The optical band gap varies between 1.843 and 1.869 eV. It is difficult to observe any systematic variation of optical band gap with change of these deposition parameters. However, the band gap of CuO films slightly increases with spin speed. The values of optical band gap with the addition of additives are tabulated in table 2. It is difficult to observe any systematic change of band gap with the addition of additives.

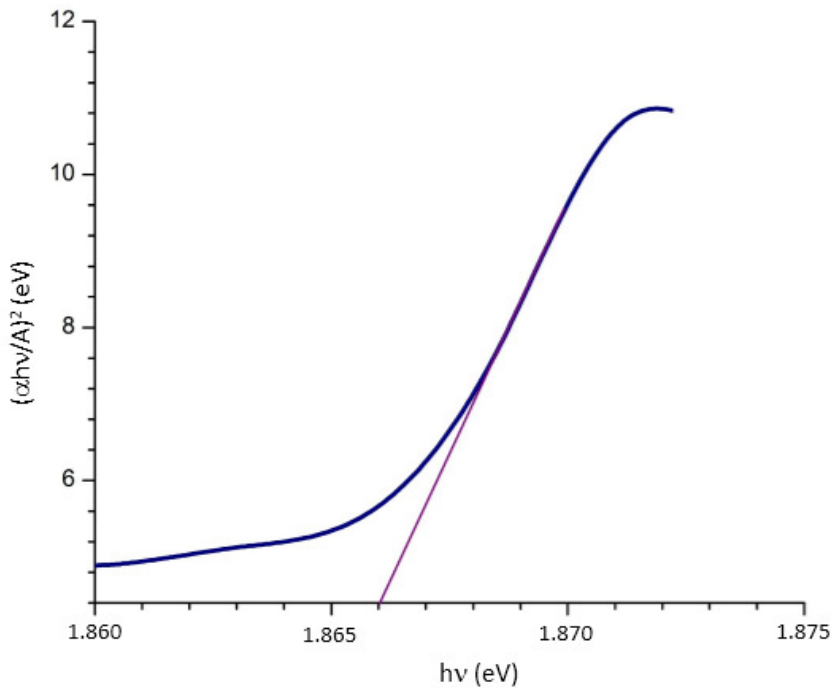

Figure 2: The graph of $(\alpha h\nu/A)^2$ vs. $h\nu$ (Photon Energy) for a thin film annealed at 350 °C.

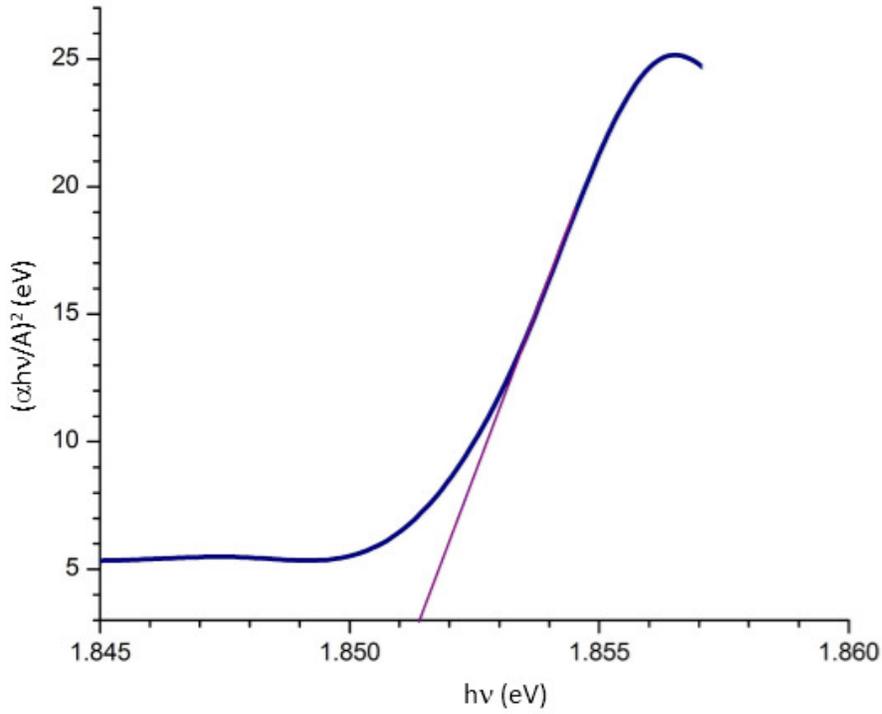

Figure 3: The graph of $(\alpha h\nu/A)^2$ vs. $h\nu$ (Photon Energy) for a thin film annealed at 450 °C.

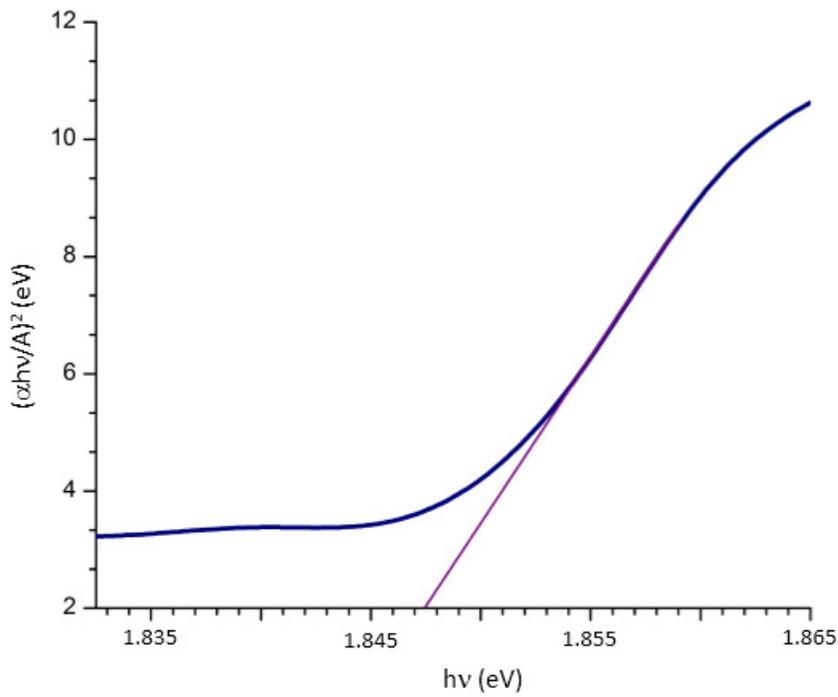

Figure 4: The graph of $(\alpha h\nu/A)^2$ vs. $h\nu$ (Photon Energy) for a thin film annealed at 550 °C.

| Spinning Rate (rpm) | Spinning Time (s) | Annealing Temperature (°C) | Annealing Time (hours) | Band Gap (eV) | Description |
|---|---|---|---|---|---|
| 1500 | 15 | 350 | 1 | 1.866 | |
| 2200 | 15 | 350 | 1 | 1.869 | |
| 1500 | 15 | 550 | 1 | 1.843 | |
| 2200 | 15 | 550 | 1 | 1.848 | |
| 1500 | 15 | 450 | 2 | 1.851 | |
| 2200 | 15 | 450 | 2 | 1.854 | |
| 1500 | 60 | 450 | 2 | 1.851 | $[Cu^{+2}]$ = 1.5 mol $dm^{-3}$ |
| 2200 | 60 | 450 | 2 | 1.855 | |
| 2400 | 15 | 450 | 2 | 1.857 | $Cu^{+2}$ : DEA = 1 : 1 |
| 2400 | 60 | 450 | 2 | 1.858 | |
| 1500 | 15 | 450 | 1 | 1.851 | |
| 2200 | 15 | 450 | 1 | 1.854 | |
| 1500 | 60 | 450 | 1 | 1.851 | |
| 2200 | 60 | 450 | 1 | 1.855 | |
| 2400 | 15 | 450 | 1 | 1.849 | |
| 2400 | 60 | 450 | 1 | 1.851 | |

Table 1: Optical band gap values of CuO films.

| Spin Rate (rpm) | Spin Time (s) | Annealing Temperature (°C) | Anneal Time (hours) | Band Gap (eV) | Description |
|---|---|---|---|---|---|
| 1500 | 30 | 550 | 1 | 1.880 | $[Cu^{+2}]$ = 1.5mol dm$^{-3}$ $Cu^{+2}$ : DEA = 1 : 1 Polyethylene Glycol: 5.0 w/w% |
| 2200 | 30 | 550 | 1 | 1.891 | |
| 1500 | 30 | 550 | 1 | 1.880 | $[Cu^{+2}]$ = 1.5 mol dm$^{-3}$ $Cu^{+2}$ : DEA = 1 : 1 Polyethylene Glycol: 10.0 w/w% |
| 2200 | 30 | 550 | 1 | 1.890 | |
| 1500 | 30 | 550 | 1 | 1.881 | $[Cu^{+2}]$ = 1.5 mol dm$^{-3}$ $Cu^{+2}$ : DEA = 1 : 1 Polyethylene Glycol: 20.0 w/w% |
| 2200 | 30 | 550 | 1 | 1.892 | |
| 1500 | 30 | 550 | 1 | 1.880 | $[Cu^{+2}]$ = 1.5 mol dm$^{-3}$ $Cu^{+2}$ : DEA = 1 : 1 Ethylene Glycol: 5.0 w/w% |
| 2200 | 30 | 550 | 1 | 1.891 | |
| 1500 | 30 | 550 | 1 | 1.882 | $[Cu^{+2}]$ = 1.5 mol dm$^{-3}$ $Cu^{+2}$ : DEA = 1 : 1.1 Ethylene Glycol: 10.0 w/w% |
| 2200 | 30 | 550 | 1 | 1.889 | |

Table 2: Optical band gap values corresponding to spin-coated CuO thin films (with additives).

Then it was experimented whether the multi-layering of cupric oxide affects the optical band gap value as tabulated in table 3. First the spin-coated thin film was annealed at 250 °C for 10 minutes. Thereafter, more layers of cupric oxide were piled on top of one another using the same annealing conditions. Finally, multi-layering was completed by being annealed it at 550 °C for one hour. According to table 3, the optical band gap decreases with the number of layers due to quantum confinement effect.

| Spin Rate (rpm) | Spin Time (s) | Annealing Temperature (°C) | Anneal Time (hours) | Band Gap (eV) | Description |
|---|---|---|---|---|---|
| 2200 | 30 | 250, 550 | 1/6, 1 | 1.831 | 3 layers |
| 2200 | 30 | 250, 550 | 1/6, 1 | 1.823 | 5 layers |

Table 3: Optical band gap values corresponding to spin-coated CuO thin films (multi-layered).

Then the effect of two additives (ethylene glycol or polyethylene glycol) on the photovoltaic properties of the cupric oxide thin film was investigated. For this analysis, CuO thin films were spin-coated on Fluorine-doped Tin Oxide (FTO) coated glass slides. Short circuit photo current density ($J_{SC}$), open circuit photo voltage ($V_{OC}$), fill factor (FF) and efficiency ($\eta$) measured using solar cell simulation system are tabulated in table 4. All these CuO films were spin coated at 2200 rpm for 30 s and annealed at 550 $^0$C for 1 hour. Photocurrent, photovoltage and efficiency increase with the addition of additives. However, the change of fill factor with the addition of additives is not systematic. The photocurrent, photovoltage and efficiency are higher with ethylene glycol compared to those with polyethylene glycol. According to our data, ethylene glycol is the better additive.

| $J_{SC}$ (mA cm$^{-2}$ × 10$^{-1}$) | $V_{OC}$ (× 10$^{-2}$ V) | FF | $\eta$ (× 10$^{-2}$) | Description |
|---|---|---|---|---|
| 1.464 | 4.446 | 0.2724 | 0.1773 | [Cu$^{+2}$] = 1.5 mol dm$^{-3}$<br>Cu$^{+2}$ : DEA = 1 : 1 |
| 7.719 | 8.972 | 0.2655 | 1.839 | [Cu$^{+2}$] = 1.5 mol dm$^{-3}$<br>Cu$^{+2}$ : DEA = 1 : 1<br>Polyethylene Glycol: 5.0 w/w% |
| 11.72 | 10.04 | 0.2719 | 3.198 | [Cu$^{+2}$] = 1.5 mol dm$^{-3}$<br>Cu$^{+2}$ : DEA = 1 : 1<br>Ethylene Glycol: 5.0 w/w% |

Table 4: Photovoltaic analysis of spin coated CuO films.

Few samples were synthesized using the sol-gel dip coating method in order to compare with the spin coating. For the sol-gel dip coated CuO thin film, the optical band gap and the photocurrent density ($J_{SC}$) were 1.821 eV and 0.135 mA cm$^{-2}$, respectively. So the results obtained for sol-gel dip coated CuO films are similar to those obtained for spin coated CuO films. However, the dip coating method is time consuming.

## 4. Conclusion:

The optical band gap decreases from 1.869 to 1.823 eV as the number of CuO layers was increased from 1 to 5. The optical band gap of CuO films slightly increases with the spin speed. By using the additives, the optical band gap of spin coated CuO films can be increased from 1.843 to 1.892 eV. All the measured optical band gap values of our CuO samples are in the range of optical band gap of bulk CuO. The photocurrent, photovoltage and the efficiency can be improved by using the additives. Ethylene glycol is a better additive compared with the polyethylene glycol. Films deposited using sol-gel dip coating exhibit the same properties as the films synthesized using spin coating. Because the same chemical reaction is used in both the sol-gel dip coating and the spin coating method, the same results were expected. Although the spin coating method could fabricate a sample in 30s, the sol-gel dip coating method needed 48 hours to synthesize a sample. So the dip coating method was found to be time consuming compared with the spin coating method.